\documentclass[conference]{IEEEtran}

\IEEEoverridecommandlockouts

\usepackage{cite}
\usepackage{amsmath,amssymb,amsfonts}
\usepackage{algorithmic}
\usepackage{graphicx}
\usepackage{textcomp}
\usepackage{xcolor}

\usepackage[T1]{fontenc}
\usepackage[utf8]{inputenc}
\usepackage{url}
\usepackage{array}
\usepackage{booktabs}
\usepackage{tabularx}
\usepackage{hyperref}
\usepackage{orcidlink}

\def\BibTeX{{\rm B\kern-.05em{\sc i\kern-.025em b}\kern-.08em
    T\kern-.1667em\lower.7ex\hbox{E}\kern-.125emX}}

\newcommand{\APG}{APG}
\newcommand{\PF}{Provenance Fidelity}
\newcommand{\kgcfr}{\texttt{kg\_cfr\_full}}

\title{Towards Mitigating Fabricated Consensus: \\ The Active Provenance Gate for \\ Multi-Agent Debate Synthesis}

\author{
\IEEEauthorblockN{Jakub Mas\l{}owski\,\orcidlink{0009-0005-4597-6335}, Jaros\l{}aw A. Chudziak\,\orcidlink{0000-0003-4534-8652}}
\IEEEauthorblockA{\textit{Institute of Computer Science} \\
\textit{Warsaw University of Technology}\\
Warsaw, Poland \\
\{jakub.maslowski2.stud, jaroslaw.chudziak\}@pw.edu.pl}
\thanks{This work has been accepted for publication at the 38th IEEE
International Conference on Tools with Artificial Intelligence
(ICTAI 2026). © 2026 IEEE. Personal use of this material is permitted.
Permission from IEEE must be obtained for all other uses, in any current
or future media, including reprinting/republishing this material for
advertising or promotional purposes, creating new collective works, for
resale or redistribution to servers or lists, or reuse of any copyrighted
component of this work in other works. The work reported in this paper
was supported by the Polish National Science Centre under grant
2024/06/Y/HS1/00197.}
}

\begin{document}

\pagestyle{empty}
\maketitle

\begin{abstract}
Large language model-based multi-agent debate (MAD) systems are being increasingly used as complex decision pipelines in distributed processes. However, their final synthesis phase still remains inadequately controlled. Even with detailed debate logs, summarizing models are prone to fabricating smoothly written debate consensus that is not grounded in the debate's history. 
To address this safety gap, this paper presents empirical research and studies if the introduction of active post-debate verification can mitigate the production of such factually unsupported summaries, while still providing valuable information. Furthermore, it is examined whether explicitly signalling divergence is preferable in the absence of a reliable compromise. The Active Provenance Gate (APG) is introduced as a post-debate verification layer that treats the source as a hard constraint, analysing the debate logs, auditing each claim, and applying self-correction. 
In crisis simulations, the self-healing mechanism more than doubles the average data Provenance Fidelity in difficult condition scenarios, before the strict gate blocks unsupported claims and generates divergence reports. In the human study, a vast majority of the users (over 75\%) preferred a report explicitly stating failure in critical scenarios, despite most of them perceiving fabricated consensus from the baseline system as more fluent. Our main contribution is the transition of data origin tracing from passive logging to active conditional blocking before publication.
\end{abstract}


\section{Introduction}

Large language models are being increasingly used as distributed decision pipelines \cite{fourney2024magentic,yao2023react} and in complex socio-educational simulations \cite{zamojska2025tacla,zamojska2025games}.
In these settings, specialized agents collaborate to collect evidence, provide feedback on each other, and collectively analyze the evidence to shape a recommendation \cite{du2024debate,fourney2024magentic}.
Multi-agent debate is therefore used to enhance the robustness of reasoning, factuality, and interpretability.
Production-grade agent stacks emphasize orchestration, role specialization, and structured traces as core components \cite{du2024debate,smit2024mad,chan2024chateval,fourney2024magentic,yao2023react}.
The usual assumption is that by reaching sufficient depth and preserving the logs, the last synthesis compresses the discussion faithfully \cite{du2024debate,chan2024chateval,smit2024mad}.

However, this assumption is a structural blind spot.
The most significant hallucination could occur not in one turn, but at the edge between deliberation and publication \cite{huang2023hallucination}.
A synthesis model optimized for coherence can dissolve an unresolved conflict into abstractions, rhetorical overlap, or softened language \cite{huang2023hallucination}.

\begin{figure}[!t]
\centering
\includegraphics[width=0.76\columnwidth]{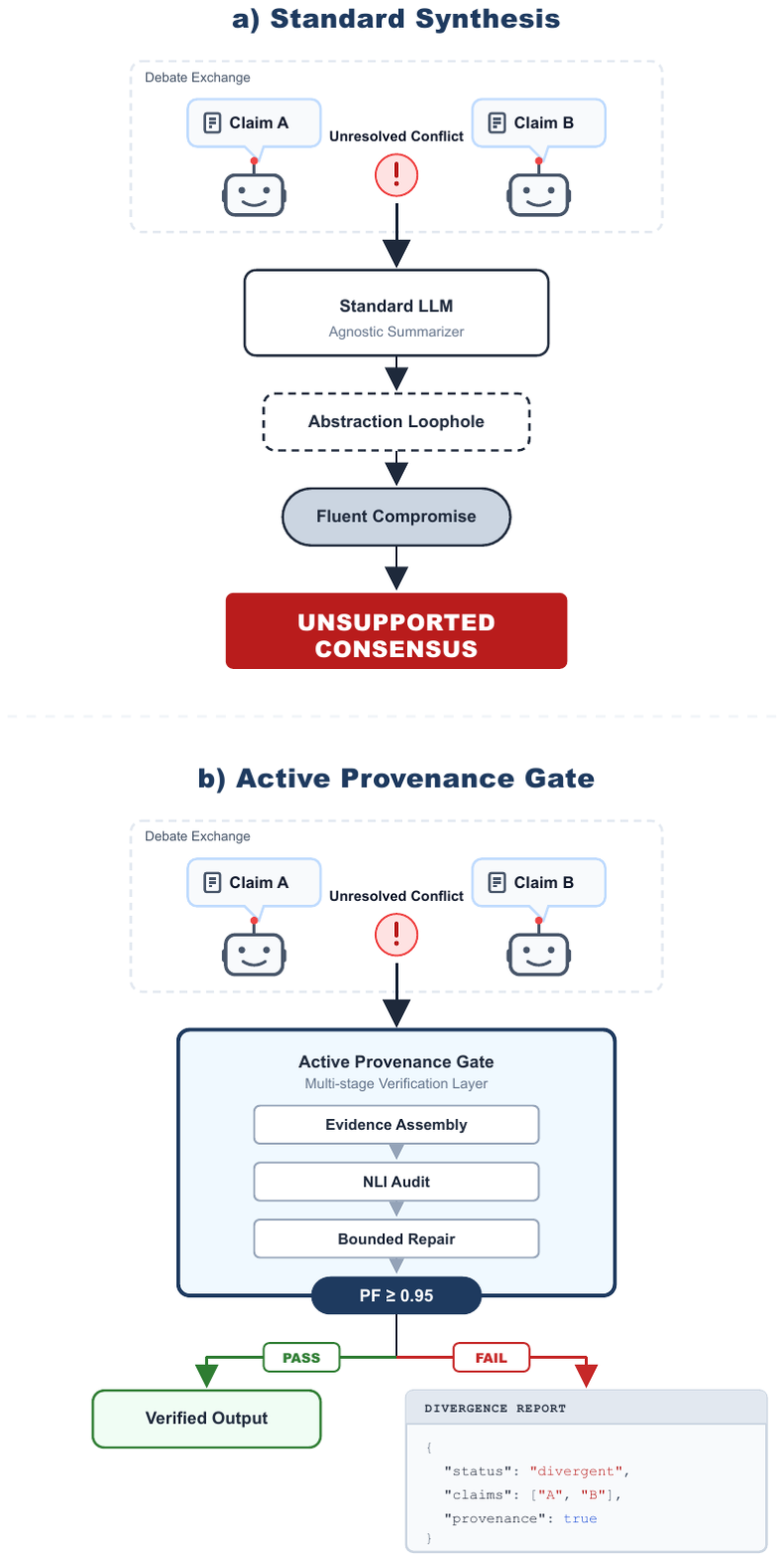}
\vspace{-3mm}
\caption{Motivating failure pattern and system intervention.
\textbf{(a)} Standard synthesis can fabricate an unsupported consensus.
\textbf{(b)} The Active Provenance Gate enforces a grounding constraint ($PF \ge 0.95$), explicitly failing safely and emitting a programmatic Divergence Report when consensus is ungrounded.}
\label{fig:hero}
\end{figure}

This problem becomes more apparent when the size of agent traces, knowledge graph axioms, and RAG system artifacts grow.
The synthesis node is more and more prone to interpolate between sources that are only partially aligned, which can cause hallucination \cite{lewis2020rag,smit2024mad,belem2025multidoc}.
This failure pattern is conceptually illustrated in the left panel (a) of Fig.~\ref{fig:hero}.

The main research question of this paper is whether the Active Provenance Gate can mitigate the generation of fabricated consensus in multi-agent debate synthesis while maintaining practical utility. This is important because high-stakes decision-support systems are less harmed by visible dissension than by subterfuge, particularly in crisis situations where consensus on a fluent agreement may miscalibrate operator trust due to automation bias.

To mitigate the generation of fabricated consensus, we thus propose the APG, conceptually illustrated in panel (b) of Fig.~\ref{fig:hero}, a layer after the debate, which treats  the debate record as a closed evidentiary world and anchors each major claim to it, blocking publication if the constraints are violated.
Instead of redesigning the debate, we govern the transition from debate to deliverable.
The system either publishes a synthesis with a verifiable provenance, or produces a Divergence Report showing conflicting stances, and lack of evidence.

Ultimately, the Active Provenance Gate (APG) makes a transition from passive tracing to a controlled publication of results. Our paper contributes to shaping provenance \cite{moreau2013provdm,groth2013provoverview} not for competitive AI, but as an operational artifact of a descriptive audit for cooperative AI \cite{souza2025provagent}.

Our research is situated at the intersection of multi-agent systems, data provenance modeling, and factual verification. Although multi-agent debates enhance the reasoning capabilities of LLMs, recent studies show that agents are prone to favoring rhetorical consistency and persuasion instead of factual grounding, which leads to a false consensus \cite{smit2024mad}. Meanwhile, existing provenance tracking frameworks, such as PROV or PROV-AGENT, treat system logs mainly as passive artifacts used for post-hoc diagnosis, not control mechanisms \cite{moreau2013provdm,groth2013provoverview,souza2025provagent}. On the other hand, popular methods of self-refine rarely enforce hard, verifiable constraints at the synthesis boundary itself \cite{madaan2023selfrefine,shinn2023reflexion}. This work attempts to address this gap by transitioning provenance from passive logging to active verification in real time.

\section{Conceptual Framework: Active Provenance Enforcement}

To mitigate the generation of fabricated consensus and unsupported synthesis in Multi-Agent Debate (MAD) architectures, our main strategy is the transition from passive logging to enforcement of active provenance.

\subsection{Principles of Active Runtime Auditing}

Current provenance-aware frameworks treat agent traces as passive diagnostic artifacts, surfaced only for post-hoc inspection \cite{moreau2013provdm,groth2013provoverview,souza2025provagent}.
We define this as \emph{Active Provenance}: enforcement of data origin tracing. It means that under active runtime auditing, each claim in the final synthesis phase must be directly connected to a particular proof from the debate log. This way, the system relies on factual grounding, not an agent's rhetoric.

This requirement becomes clearly visible in knowledge-grounded architectures, where agents retrieve dense Knowledge Graph (KG) axioms \cite{yao2023react} and Retrieval-Augmented Generation (RAG) axioms \cite{lewis2020rag}. Paradoxically, such a large amount of data facilitates the models' creation of false connections between facts. When LLMs are fed with too much conflicting information, they start to generate answers that sound reasonable, but do not have source grounding \cite{huang2023hallucination}. Therefore, active runtime auditing is a structural requirement for maintaining fidelity, not just an optional feature.

\subsection{Research-by-Construction Methodology}

This study adopts a research-by-construction approach, utilizing a controlled Adversarial Debate Testbed (Phases 1--3).
Rather than selecting a benchmark post-hoc, we target unsupported consensus by engineering a setting that yields high-fidelity traces under systematic epistemic stress.

The testbed is a simulation of a crisis-management committee with a fixed budget.
Epistemic shocks (e.g., conflicting telemetry) are used to cause logical deadlocks.
Restricting debate exclusively to knowledge-grounded traces reduces ungrounded agent drift.
Consequently, unsupported syntheses can be directly attributed to the publication-stage synthesis module, making the failure mode visible and interruptible.
\subsection{Formalization of Provenance Fidelity (PF)}

To measure evidential grounding, we formalize Provenance Fidelity (PF) as the ratio of supported semantic units (evaluated at the sentence level) to total output sentences.
Let $C$ be the generated sentences, and $A(c) \in \{0, 1\}$ represent strict NLI entailment with the corpus:

\begin{equation}
\mathrm{PF} = \frac{\sum_{c \in C} A(c)}{|C|}
\label{eq:pf}
\end{equation}
\vspace{-3mm}

A PF score of $1.0$ indicates perfect traceability;
lower scores reflect ungrounded interpolation.
Operationally, the \textsc{VALIDATOR NODE} (Gemini 3.1 Pro\footnote{Model versions correspond to the state as of May 2026.}) acts as a strict Natural Language Inference (NLI) engine, mapping claims against public transcripts and axioms to provide verifiable anchoring \cite{bowman2015snli,tang2024minicheck,shinn2023reflexion}.

\section{System Architecture: The Active Provenance Gate}

\begin{figure}[!t]
\centering

\includegraphics[width=0.75\columnwidth, viewport=40 130 572 720, clip=true, keepaspectratio]{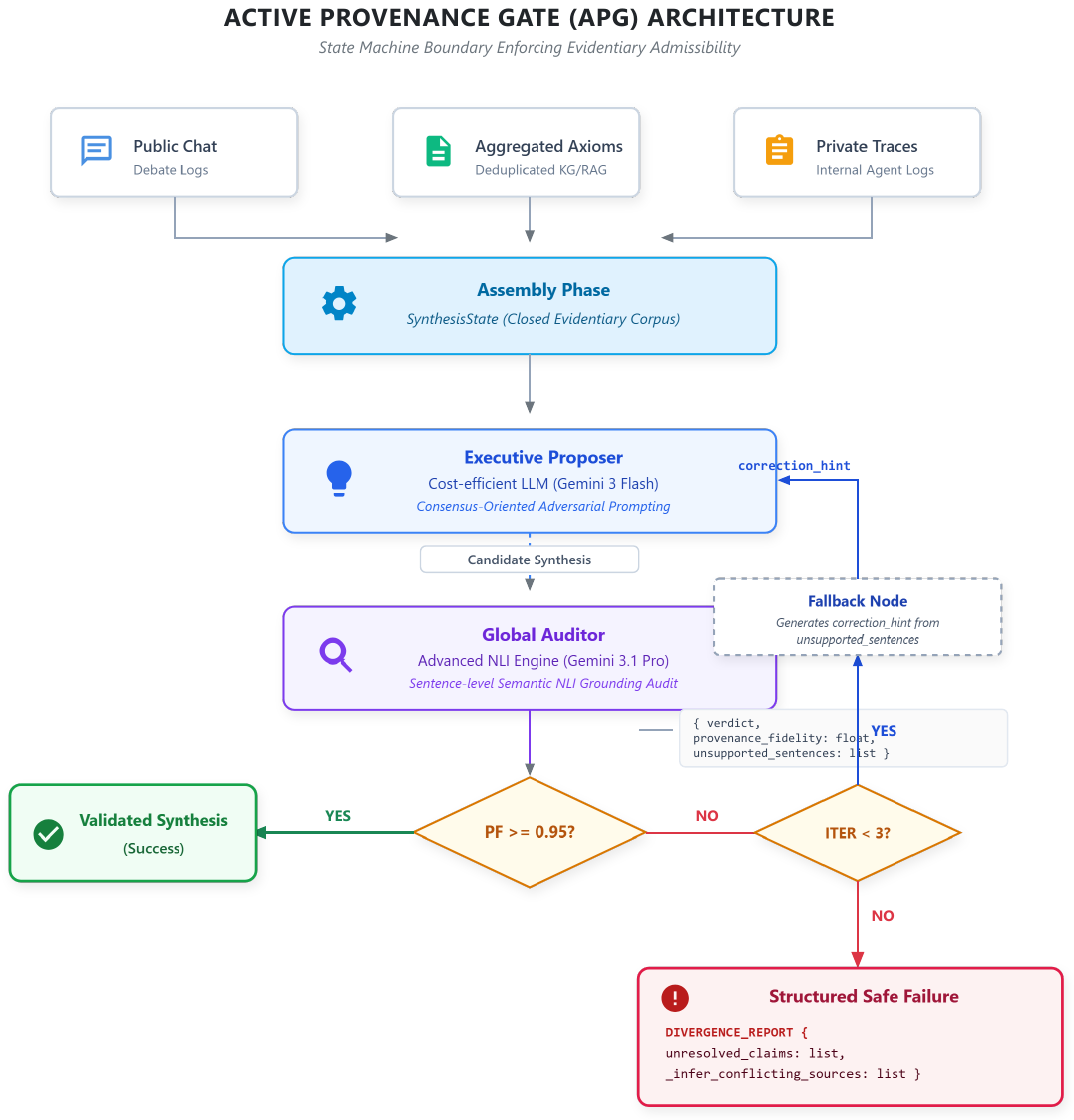}
\vspace{-3mm}
\caption{State machine architecture of the Active Provenance Gate. The system aggregates debate artifacts into a closed evidentiary corpus, decoupling generation (Executive Proposer) from verification (Global Auditor). A hard computational gate ($PF \ge 0.95$) enforces NLI auditing. If a synthesis is ungrounded, bounded self-healing is triggered ($ITER < 3$); exhausted budgets trigger a hard exit via a programmatic Divergence Report.}
\label{fig:apg}
\end{figure}

The Active Provenance Gate (\APG{}) is a particular loop (Phase 4) inserted between the multi-agent debate and publication.
It is implemented as a deterministic state machine using \texttt{LangGraph} (see Fig.~\ref{fig:apg}).
It acts as a post-debate hard gate wrapping the synthesis boundary with four functional components enforcing evidentiary admissibility, NLI auditing, bounded self-healing, and explicit failure signalling.

\subsection{Evidentiary Corpus and Assembly Phase}

At the end of Phase 3, the system merges all session artifacts into a single \texttt{SynthesisState} object.
This Assembly phase builds a closed evidentiary world aggregating: (i) the public chat transcript, (ii) deduplicated KG/RAG documents extracted via \texttt{collect\_aggregated\_axioms}, and (iii) private traces collected with internal strategic intents via \texttt{collect\_private\_traces}.
These artifacts are not meant to be used as a loose context window, but rather are normalized to a bounded corpus with stable addresses \cite{moreau2013provdm,groth2013provoverview,souza2025provagent}.

After the corpus is frozen, the Proposer works in a strict closed-world regime. Legitimate synthesis operations can only involve compression, conflict exposure, and logical deduction. Since hallucinations frequently appear as believable bridges between partially compatible fragments, converting the context into a strictly bounded payload allows the \APG{} to turn these ungrounded bridges into auditable violations \cite{huang2023hallucination,tang2024minicheck}.

\subsection{The Verification Gate: Adversarial Constraints and NLI Auditing}

The essence of the \APG{} is the interaction between a heterogeneous Proposer--Validator pair. To optimize cost and safety, the architecture is based on a cost-effective LLM as the \emph{Executive Proposer} \cite{fourney2024magentic} and an advanced reasoning model as the \emph{Global Auditor} \cite{tang2024minicheck}.

The Proposer generates a candidate synthesis under a consensus-oriented prompting strategy.
The system does not expose an explicit no-consensus response mode at the proposer stage, which reflects decision-support settings where operational recommendations are supposed to be made under time pressure.
Under epistemic shock, this pressure causes unsupported compromise formation that can be empirically observed in our testbed \cite{huang2023hallucination}.

The Auditor is a strict NLI engine, following the broader use of entailment and fact-verification datasets for judging whether generated claims are supported by evidence \cite{thorne2018fever}.
When asked to rate the support for each sentence, the Auditor carries out semantic segmentation implicitly at inference time, while sentence boundaries for post-hoc \PF{} computation are determined deterministically using standard regex-based delimiters in a list-like manner.
It executes NLI entailment checks per sentence and returns a strict JSON payload containing a \texttt{verdict}, a \texttt{provenance\_fidelity} ratio, and \texttt{unsupported\_sentences}.
If the score falls below the programmatic hard gate (e.g., $\mathrm{PF} = 0.95$), the \APG{} will overrule any positive local verdict and reject it, capturing the exact ungrounded claims for repair \cite{tang2024minicheck}.

\subsection{Bounded Self-Healing and Divergence Reporting}

When validation fails, the \textsc{FALLBACK NODE} starts a bounded self-correction loop.
It uses the unsupported sentences to generate a hint that instructs the \textsc{PROPOSER NODE} (Gemini 3 Flash) to eliminate or neutralize these assertions.
To ensure that there are no infinite hallucination loops in unresolvable conflicts, there is a hard limit on the number of retries ($\mathrm{MAX\_SYNTHESIS\_ITER} = 3$) \cite{madaan2023selfrefine,shinn2023reflexion}.

Failing to reach the threshold ($\mathrm{PF} \ge 0.95$) on these attempts, the \APG{} will override any positive LLM verdict and force a hard exit.
Instead of outputting an ungrounded compromise, it defaults to explicitly signaling failure, outputting a JSON \texttt{DIVERGENCE\_REPORT}.
This gives the actionable intelligence by listing \texttt{unresolved\_claims} and using \texttt{\_infer\_conflicting\_sources} to pinpoint traces that lead to deadlock.

% ==========================================
% NOWA SEKCJA 4: Experimental Setup
% ==========================================
\section{Experimental Setup}

We assess \APG{} on three aspects: synthesis grounding under adversarial conflict, calibrated user trust, and operational verification cost. Throughout all the experiments, models were configured with Temp: 0.5 and Max Tokens: 8192. 

The grounding experiment is based on the \kgcfr{} corpus, which is fully integrated into our open-source Resilient MAS Framework \cite{resilientmas2026}. This comprehensive testbed contains 90 knowledge-annotated trajectories, full prompt templates, and the complete verification codebase to ensure full reproducibility \cite{kgcfrfull2026,maslowski2026decoupling}. We retain the same consensus-oriented prompting regime to model decision-support settings that punish indecision.

% \subsection{Human-Centric Study Design}

To evaluate the impact of reporting divergence on decision-making, we conducted a blind A/B study ($N=33$; see repository \cite{kgcfrfull2026} for survey data). Participants were presented with raw debate traces and two candidate syntheses: (A) a fluent Baseline and (B) our \APG{} report (including Divergence Reports for $S_{026}$).

Candidates were recruited by convenience sampling, where the majority were active 
students or graduates of higher education institutions, predominantly from STEM and 
technical disciplines. 21.2\% of the group declared high literacy in using LLMs, 
while the other two groups of intermediate (48.5\%) and elementary (30.3\%) 
knowledge provided a natural diversity of user competence.

% \subsection{Model Ablation and Operational Profiling}

Finally, we discuss the verification loop's architecture and a targeted 30-run ablation study addressing controllability and computational overhead. We compare a homogeneous pipeline---where the Proposer and Validator use the same highly efficient model class---with a mixed setup pairing the efficient Proposer with a high-reasoning expert model focused on NLI \cite{tang2024minicheck}.

This tests the hypothesis that verification disproportionately benefits from advanced reasoning, evaluating whether symmetrical setups fail to capture their own sophisticated hallucinations. We also examine the interaction between validator strictness and system divergence: a permissible threshold increases the risk of ungrounded synthesis, while an aggressive one is at risk of inhibiting beneficial syntheses \cite{tang2024minicheck}.

% ==========================================
% 5: Results
% ==========================================
\section{Results}

Our analysis describes the trade-off between synthesis grounding and operational divergence in multi-agent deliberations. By isolating the publication boundary via the \APG{}, results show that passive evidentiary density cannot prevent LLM hallucination under epistemic stress \cite{huang2023hallucination}. Simultaneously, a self-healing mechanism \cite{madaan2023selfrefine,shinn2023reflexion}, along with the use of low-latency models \cite{tang2024minicheck}, allows for maintaining low computational costs while keeping the system's operation efficient.

\begin{figure}[!t]
    \centering
    \includegraphics[width=0.78\columnwidth,keepaspectratio]{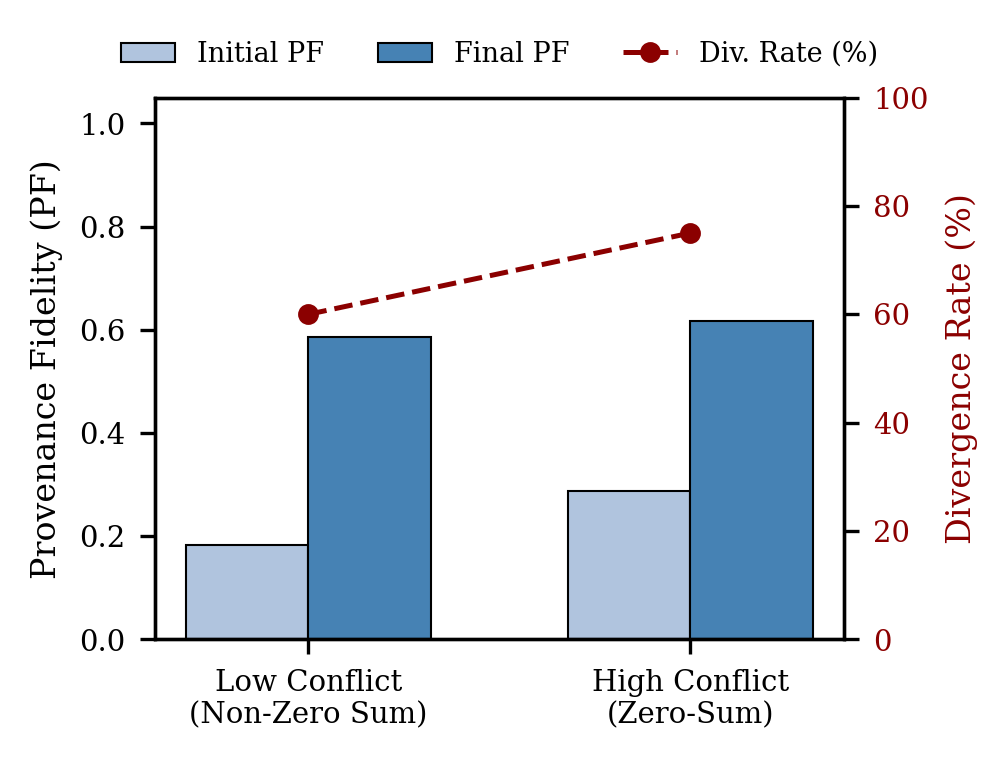}
    \vspace{-3mm}
    \caption{Synthesis grounding under epistemic shock.}
    \label{fig:shock}
\end{figure}

\begin{figure}[t]
    \centering
    \includegraphics[width=0.9\linewidth]{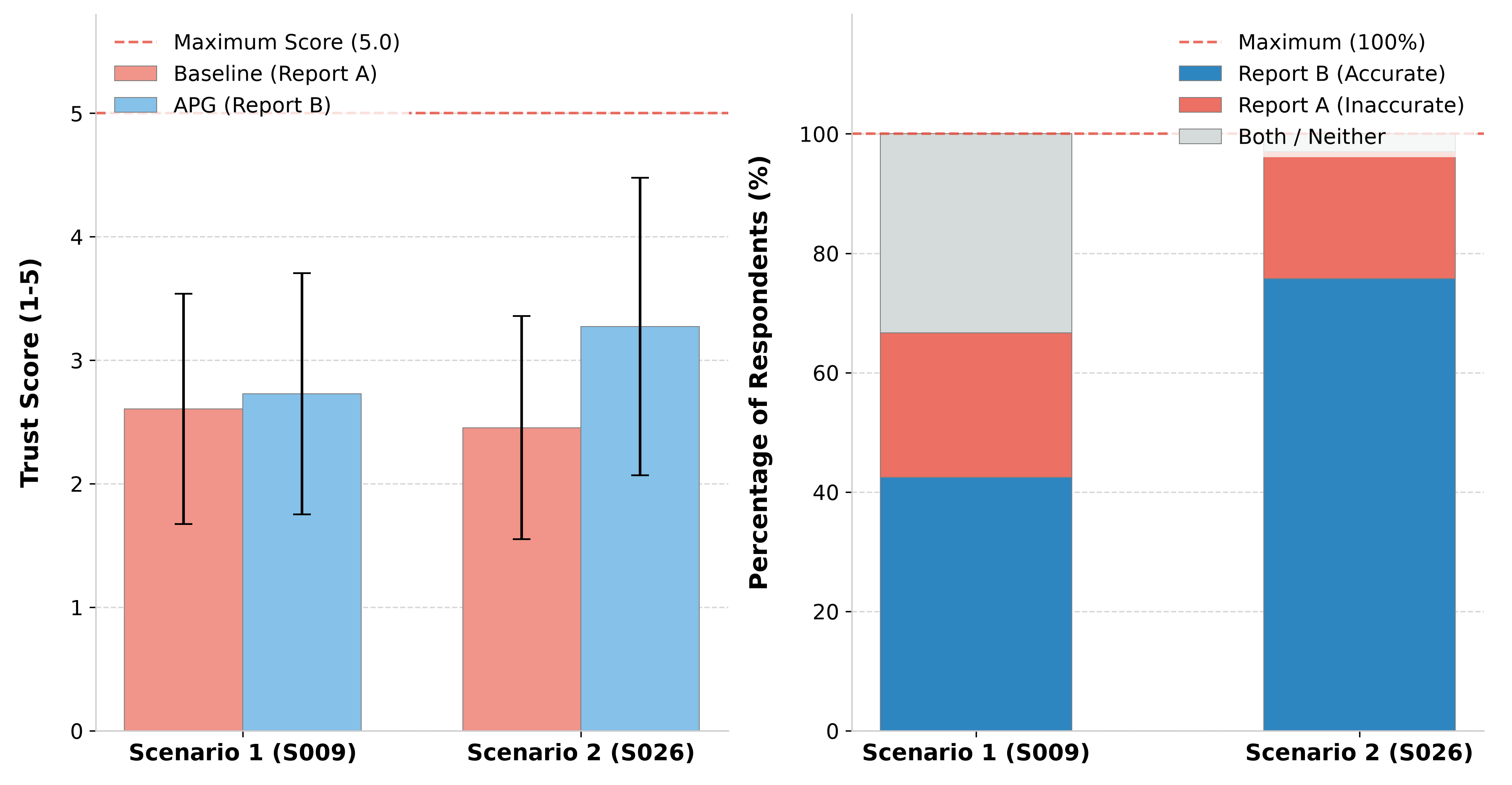}
    \vspace{-3mm}
    \caption{Calibrated Trust and Groundedness Perception}
    \label{fig:human_study}
\end{figure}

% [WCZEŚNIEJ]: Stara podsekcja 6.1 (bez zmian)
\subsection{Impact of Active Auditing on Synthesis Grounding}

Across the entire knowledge-grounded dataset, the baseline synthesis had very high hallucination rates, resulting in critically low initial Provenance Fidelity scores (0.288 under high conflict, 0.183 under low).
Even with perfectly grounded RAG/KG traces, unconstrained synthesis nodes attempt to compress incompatible axioms into connective logic.
Denser evidence combinatorially extends the unsupported interpolation surface \cite{lewis2020rag,huang2023hallucination}. Active auditing significantly reduced this failure rate through bounded self-healing, raising the average terminal \PF{} (before the final gate evaluation) to 0.617 and 0.586, respectively. Fig.~\ref{fig:shock} illustrates the impact of epistemic shock on synthesis grounding. While initial baseline fidelity (\PF{}) remains low, the bounded self-healing loop elevates the Terminal PF. The divergence rate captures the percentage of these terminal outputs that still failed to meet the strict $0.95$ publication threshold, triggering a programmatic Divergence Report as conflict intensifies.

Crucially, the architecture exhibits sensitivity to epistemic shock rather than arbitrary rejection.
Under severe conflict enforcing zero-sum deadlocks, the gate yielded a 75.0\% divergence rate, emitting a Divergence Report because the terminal synthesis failed to meet the strict publication threshold ($PF\ge0.95$).
Conversely, this divergence rate fell to 60.0\% in low-conflict environments. Consequently, the \APG{} enforces a verifiable grounding floor.
A conventional summarizer works as intended under mild conflict, yet becomes unreliable when stakes are highest and pressure to force closure is strongest.
Due to imposing the hard verification boundary, the system does not silently 
ignore factually unsupported consensus, instead 
actionably rejecting such a response.

% [WCZEŚNIEJ]: Stara podsekcja 6.3 (teraz jako 6.2)
\subsection{Operational Efficiency and Parameter Sensitivity}

Validator ablation exposes a vulnerability in symmetrical deployments.
Using Gemini 3 Flash symmetrically as both generator and NLI auditor wrongly validated structurally unsafe compromises 85.6\% of the time \cite{tang2024minicheck}.
Switching to an asymmetric configuration (deploying Gemini 3.1 Pro only as \textsc{VALIDATOR NODE}) significantly reduced this failure pattern, downgraded identical texts to a \PF{} of 0.300, and rejected 63.3\% of irreconcilable outputs.
This means verification requires distinct model capacities to maintain adversarial pressure.

A preliminary sensitivity sweep ($n=10$ per threshold) suggested that $\mathrm{PF}=0.95$ provided a practical trade-off compared to the more permissive ($\mathrm{PF} = 0.77$) and the stiffer ($\mathrm{PF}=0.99$) options.
Furthermore, the $\mathrm{MAX\_SYNTHESIS\_ITER}=3$ limit is empirically justified: according to system logs, sessions that fail after two correction hints usually get stuck in oscillatory loops, making further retries computationally inefficient \cite{fourney2024magentic}.

% [WCZEŚNIEJ]: Druga połowa starej podsekcji 5.2 (Twarde wyniki z "Human-Centric Study")
\subsection{Calibrated Trust Dynamics}

As shown in Fig.~\ref{fig:human_study}, results suggest a persistent tendency toward unsupported consensus. While the Baseline was rated more ``fluent and diplomatic'' by 66.7\% of respondents, users were highly sensitive to grounding failures. In the high-shock scenario ($S_{026}$), 75.8\% of participants identified the \APG{} output as more faithful to the evidence, effectively piercing through the seductive fluency of the fabricated consensus.

Crucially, we measured \textit{Decision Utility} via calibrated trust on a 5-point Likert scale.
In the energy allocation scenario ($S_{009}$), trust levels were comparable. However, in the zero-sum stalemate ($S_{026}$), the \APG's explicit divergence report (Report B) achieved higher trust ($M=3.27, SD=1.21$) compared to the Baseline's fabricated compromise ($M=2.45, SD=0.90$). A paired t-test indicated this difference as statistically significant ($t(32) = 3.12, p = 0.0039$). This supports our hypothesis: users prefer an explicit admission of system failure over a fluent but ungrounded hallucination of consensus.

% ==========================================
% Discussion and Evaluation
% ==========================================
\section{Discussion and Evaluation}

The bounded regeneration loop exhibits very efficient self-healing. Recovered sessions showed 97.3\% text length retention with a small semantic overlap between the initial drafts and the final draft. From an analytical point of view, the system is able to achieve compliance not via destructive truncation, but by restructuring directives around verified axioms \cite{madaan2023selfrefine,shinn2023reflexion}. Grounded synthesis and utility are jointly recoverable, and restriction of this ungrounded synthesis reveals the models' dependence on rhetorical evasion. After the rhetoric has been removed, the system defaulted to creating operational data, rephrasing fluency as a risk factor \cite{huang2023hallucination}. Conversely, structured Divergence Reports identified deliberative deadlocks effectively. Therefore, in the event of a serious conflict, the explicit failure reporting paradigm was a significant factor in enhancing calibrated trust.

However, our research is limited by several factors. Our evaluation is based on synthetic simulations, and hard-coded shocks. The \APG{} is based on a closed-world assumption, evaluating provenance against the aggregated evidentiary corpus, instead of against external reality \cite{moreau2013provdm,groth2013provoverview}. The proxy study ($N=33$) supports trust calibration; however, as the lack of division between domain experts and crowd-sourced operators warrants a cautious interpretation of calibrated trust, an important next step would be to test generalizability to professional operators under time pressure in real-time. Finally, zero-shot LLM validation for NLI auditing is susceptible to linguistic biases, verbosity biases, and prompt sensitivity \cite{tang2024minicheck,wang2024factualitysurvey}. Although model heterogeneity reduces the vulnerability of poor self-evaluation, the intrinsic subjectivity of the verification process remains a defined system boundary. What is important, using an LLM auditor as an operational reference point creates a risk of metric circularity. Comparing it with external NLI benchmarks or human annotations remains a crucial verification step.

% ==========================================
% Conclusions
% ==========================================

\section{Conclusions}

Multi-agent debate pipelines are becoming more compact, but as distributed deliberations turn into consequential decisions, the publication boundary between debate and deliverable remains weakly governed.
We observed that fluency-optimized synthesis models systematically make use of this gap, turning irreconcilable grounded positions into fabricated consensus---a serious failure mode that is not apparent to passive provenance logging.

Our main contribution is the \APG{}: a post-debate enforcement layer on top of the deliberation trace as a strictly closed evidentiary world.
By engaging in active auditing of candidate syntheses, the \APG{} enforces a publication restriction through a rigorous evidential fidelity threshold ($\mathrm{PF} \ge 0.95$).
Across 90 adversarial runs, we showed that this architecture increases the Provenance Fidelity floor, forces bounded repair, and provides a structured divergence mechanism when grounded agreement is not reachable in our testbed.
Consequently, we present the potential to convert a passive diagnostic into an active, runtime publication constraint.

Future work will involve expanding active provenance enforcement directly into the real-time deliberation loop, rather than solely at the synthesis boundary.
We also plan to deploy dynamically calibrated claim-level judges working in tandem, for higher validation quality, and design interactive Divergence Reports for operator-in-the-loop arbitration.

\bibliographystyle{IEEEtran}
\bibliography{references}

\begin{thebibliography}{10}
\providecommand{\url}[1]{#1}
\csname url@samestyle\endcsname
\providecommand{\newblock}{\relax}
\providecommand{\bibinfo}[2]{#2}
\providecommand{\BIBentrySTDinterwordspacing}{\spaceskip=0pt\relax}
\providecommand{\BIBentryALTinterwordstretchfactor}{4}
\providecommand{\BIBentryALTinterwordspacing}{\spaceskip=\fontdimen2\font plus
\BIBentryALTinterwordstretchfactor\fontdimen3\font minus \fontdimen4\font\relax}
\providecommand{\BIBforeignlanguage}[2]{{%
\expandafter\ifx\csname l@#1\endcsname\relax
\typeout{** WARNING: IEEEtran.bst: No hyphenation pattern has been}%
\typeout{** loaded for the language `#1'. Using the pattern for}%
\typeout{** the default language instead.}%
\else
\language=\csname l@#1\endcsname
\fi
#2}}
\providecommand{\BIBdecl}{\relax}
\BIBdecl

\bibitem{fourney2024magentic}
A.~Fourney \emph{et~al.}, ``{Magentic-One}: A generalist multi-agent system for solving complex tasks,'' \emph{arXiv preprint arXiv:2411.04468}, 2024.

\bibitem{yao2023react}
S.~Yao \emph{et~al.}, ``{ReAct}: Synergizing reasoning and acting in language models,'' in \emph{Proc. ICLR}, 2023.

\bibitem{zamojska2025tacla}
M.~Zamojska and J.~A. Chudziak, ``{TACLA}: An {LLM}-based multi-agent tool for transactional analysis training in education,'' in \emph{Proc. ICTAI}, 2025, pp. 313--320.

\bibitem{zamojska2025games}
------, ``Games agents play: Towards transactional analysis in {LLM}-based multi-agent systems,'' in \emph{Proc. CogSci}, 2025.

\bibitem{du2024debate}
Y.~Du \emph{et~al.}, ``Improving factuality and reasoning in language models through multiagent debate,'' in \emph{Proc. ICML}, 2024.

\bibitem{smit2024mad}
A.~P. Smit \emph{et~al.}, ``Should we be going {MAD}? a look at multi-agent debate strategies for {LLMs},'' in \emph{Proc. ICML}, 2024, pp. 45\,883--45\,905.

\bibitem{chan2024chateval}
C.-M. Chan \emph{et~al.}, ``{ChatEval}: Towards better {LLM}-based evaluators through multi-agent debate,'' in \emph{Proc. ICLR}, 2024.

\bibitem{huang2023hallucination}
L.~Huang \emph{et~al.}, ``A survey on hallucination in large language models,'' \emph{ACM TOIS}, 2024.

\bibitem{lewis2020rag}
P.~Lewis \emph{et~al.}, ``Retrieval-augmented generation for knowledge-intensive {NLP} tasks,'' in \emph{Proc. NeurIPS}, vol.~33, 2020, pp. 9459--9474.

\bibitem{belem2025multidoc}
C.~G. Bel{\'e}m \emph{et~al.}, ``From single to multi: How {LLMs} hallucinate in multi-document summarization,'' in \emph{Findings of NAACL}, 2025.

\bibitem{moreau2013provdm}
{W3C Provenance Working Group}, ``{PROV-DM}: The {PROV} data model,'' 2013.

\bibitem{groth2013provoverview}
P.~Groth \emph{et~al.}, ``{PROV-Overview}: An overview of the {PROV} family of documents,'' 2013.

\bibitem{souza2025provagent}
R.~Souza \emph{et~al.}, ``{PROV-AGENT}: Unified provenance for tracking {AI} agent interactions in agentic workflows,'' in \emph{Proc. eScience}, 2025.

\bibitem{madaan2023selfrefine}
A.~Madaan \emph{et~al.}, ``Self-refine: Iterative refinement with self-feedback,'' in \emph{Proc. NeurIPS}, vol.~36, 2023.

\bibitem{shinn2023reflexion}
N.~Shinn \emph{et~al.}, ``Reflexion: language agents with verbal reinforcement learning,'' \emph{Adv. Neural Inf. Process. Syst.}, vol.~36, pp. 8634--8652, 2023.

\bibitem{bowman2015snli}
S.~R. Bowman \emph{et~al.}, ``A large annotated corpus for learning natural language inference,'' in \emph{Proc. EMNLP}, 2015, pp. 632--642.

\bibitem{tang2024minicheck}
L.~Tang \emph{et~al.}, ``{MiniCheck}: Efficient fact-checking of {LLMs} on grounding documents,'' in \emph{Proc. EMNLP}, 2024, pp. 8818--8847.

\bibitem{thorne2018fever}
J.~Thorne \emph{et~al.}, ``{FEVER}: a large-scale dataset for fact extraction and {VERification},'' in \emph{Proc. NAACL}, 2018, pp. 809--819.

\bibitem{resilientmas2026}
J.~Mas\l{}owski, ``Resilient {MAS} framework: A testbed for ethical {AI} in crisis scenarios,'' 2026, \url{https://github.com/m-Jakub/resilient-mas-framework}.

\bibitem{kgcfrfull2026}
------, ``Multi-agent debate corpus ({kg\_cfr\_full}) and {APG} reproducibility bundle,'' 2026, \url{https://github.com/m-Jakub/resilient-mas-framework/tree/main/paper_supplementary/article_3_apg}.

\bibitem{maslowski2026decoupling}
J.~Mas{\l}owski and J.~A. Chudziak, ``Decoupling thought from speech: Knowledge-grounded counterfactual reasoning for resilient multi-agent argumentation,'' in \emph{Proc. KES}, 2026.

\bibitem{wang2024factualitysurvey}
Y.~Wang \emph{et~al.}, ``Factuality of large language models: A survey,'' in \emph{Proc. EMNLP}, 2024, pp. 19\,519--19\,529.

\end{thebibliography}
\end{document}